\documentclass[10pt,prl,twocolumn,showpacs]{revtex4}

\usepackage{graphicx}
\usepackage{amsmath}
\usepackage{epsfig}
\usepackage{epstopdf}

\newcommand{\rmq}[1]{}
\newcommand{\old}[1]{}

\newcommand{\be}{\begin{equation}}
\newcommand{\ba}{\begin{eqnarray}}
\newcommand{\ea}{\end{eqnarray}}
\newcommand{\ee}{\end{equation}}

\newcommand{\Cal}[1]{{\cal #1}}
\newcommand{\p}{\partial}

\newcommand{\vm}[1]{\left\langle #1 \right\rangle}
\newcommand{\XXX}[1]{}
\newcommand{\II}{I \! \! I}
\newcommand{\bea}{\begin{eqnarray}}
\newcommand{\eea}{\end{eqnarray}}
\newcommand{\prt}[1]{\left( #1\right)}
\newcommand{\ir}{{\tiny  \mbox{IR}}}

\begin{document}
\bibliographystyle{apsrev}


\title{Exact  low temperature results for transport properties of 
the interacting resonant level model}

\author{E.~Boulat$^{1,2}$, H.~Saleur$^{2,3}$}
\affiliation{${}^1$Laboratoire MPQ, Universit\'e Paris Diderot, 75205 Paris Cedex 13}
\affiliation{${}^2$Service de Physique Th\'eorique, CEN Saclay, Gif 
Sur Yvette, F-91191}
\affiliation{${}^3$Department of Physics,
University of Southern California, Los Angeles, CA 90089-0484}


\date{\today}

\begin{abstract}
Using  conformal field theory and integrability ideas, we give a full characterization of the low temperature regime of the anisotropic  interacting resonant level (IRLM)  model. We determine the low temperature corrections to the linear conductance exactly up to the 6th order. We show that the structure displays 'Coulomb deblocking' at resonance, i.e., a strong impurity-wire capacitive coupling enhances the conductance at low temperature.
\end{abstract}

\maketitle

Transport properties in quantum impurity problems have become central to experimental nanophysics \cite{rev}.  Their theoretical study involves strong interactions, often out of equilibrium, and thus  presents considerable  difficulties. In many cases, it is possible to  map these systems to simple one dimensional models, which can be solved  by the Bethe ansatz (BA) in equilibrium. Extending the BA out of equilibrium is the next step forward. This has been  accomplished in the problem of edge state tunneling in the fractional quantum Hall effect (where coupling to the reservoirs was particularly simple)  \cite{FLS},\cite{BLZ}, but major obstacles remain in most other cases \cite{KLS}. 
Recently, an exciting new ``open Bethe ansatz'' has  been proposed \cite{Natan}, which might well be a breakthrough, although subtle issues remain open - related to universality and the treatment of boundary conditions around the impurity out of equilibrium: see \cite{Benjamin} for recent progress in this direction.

It is one of the major difficulties of this field that so few methods are available to investigate strongly correlated regimes that it is often not possible to assert the validity of results such as those in \cite{FLS} or \cite{Natan} (see eg \cite{Borda}).   Our goal in this paper is to report on a method to tackle low temperature properties in the IRLM which gives highly non perturbative results, and apart from 
its own usefulness, provides results that can  be used as  benchmarks. Another, different perturbative approach has been proposed recently in \cite{Benjamin}.

The IRLM  model  describes  a resonant level coupled via tunneling junctions to two baths of spinless electrons, with which there is also a Coulomb interaction \cite{Natan,Andrei}. 
After the usual expansion in angular modes, unfolding and linearizing near the Fermi surface, one ends up with a hamiltonian 
$H = H_0 + H_B$ where 
$H_{0}= -i\sum_{a=1,2}\int dx\, \psi^\dagger_a\p_x\psi^{}_a$ 
is the free hamiltonian describing two infinite right moving Fermi wires, and tunneling occurs through the impurity term:
\be
\begin{array}{c}
H_B = \left(\gamma_{1}\psi^{}_1(0)+\gamma_{2}\psi^{}_2(0)\right)\,d^\dagger +
\mbox{h.c.}
\vspace{0.1cm}\\
+\;U \;
\big(\psi^\dagger_1\psi^{}_1(0)+\psi^\dagger_2\psi^{}_2(0)\big) \, \big(d^\dagger d-\textstyle\frac{1}{2}\big) + \epsilon_{d}~d^\dagger d
\end{array}
\label{hamiltonian}
\ee
In the following, it will be convenient to use the langage of the Kondo model, which is unitarily related to the IRLM \cite{wiegmann78}, and to introduce spin $1/2$ operators to represent the impurity: $d^\dagger=\eta S^+$, $d^\dagger d=S^z+\frac{1}{2}$ ($\eta$ is a Majorana fermion).
The parameters $\gamma_{1,2}$ (which can be taken real) are parametrized as $\gamma_{1}+i\gamma_{2}=\gamma\sqrt{2}e^{i\theta/2}$ ; $\theta$  encodes anisotropy in the tunneling process. 
Note the presence of the important interaction term, a capacitive coupling $U$ between the impurity and the wires. When the on-site chemical potential $\epsilon_{d}$ vanishes, the impurity is at resonance (in an actual experiment this would require adjusting the local grid potential to some value $V_g(U)$).

We shall be concerned with the conductance of the structure when a voltage $V$, that couples as $H_V=\sum_{a=1,2}\int dx\, (\psi^\dagger_1\psi^{}_1-\psi^\dagger_2\psi^{}_2)$, is applied across the impurity.
The standard approach to this model is to form the combinations 
$\psi^{}_{+/-}=\frac{1}{\gamma\sqrt{2}}
\big(\gamma_{1/2}^{}\psi^{}_1\pm\gamma_{2/1}^{}\psi^{}_2\big)$,
which lead to  decoupling into   two independent sectors, where $H$ 
can be  diagonalized using a straightforward BA \cite{Wiegmann}. 
Using Friedel's sum rule the linear conductance $G={dI\over dV}\big|_{V=0}$ is given at temperature $T=0$ by:
\begin{equation}
G_\ir = \textstyle{\frac{e^{2}}{ h}} \;\sin^{2}\theta \;\sin^{2}(\pi n_d)
\label{gfriedel}
\end{equation}
\vspace{-0.2cm}

The term $\sin^2\theta=\frac{2\gamma_1\gamma_2}{\gamma_1^2+\gamma_2^2}$ is the familiar tunneling anisotropy prefactor ; the impurity filling $n_d=\vm{d^\dagger d}$ can be extracted in 'closed form' at zero temperature using BA in the $\psi^{}_\pm$ basis \cite{ponomarenko93}. 
However, going beyond formula (\ref{gfriedel}), which is valid only at $V=T=0$ is a very difficult task within this BA. The problem is that  one is typically interested in situations where the two wires are at different chemical potentials, an ensemble very difficult to represent in the $\psi^{}_{\pm}$ basis.
We will show in this paper that in the field theory limit, several important results can nevertheless be obtained by exploiting hidden symmetries. 
We will present the boundary conditions characterizing the zero temperature or infrared (IR) fixed point by making use of an underlying SU(2) structure.
This gives an alternative, straightforward way to obtain the IR conductance (\ref{gfriedel}) but also allows for the setting up of a convenient formalism (perturbed boundary conformal field theory) for systematically obtaining the low $T$ corrections to conductance (formulae (\ref{girpert},\ref{corrg})).

Our results rely on two equivalent representations of (\ref{hamiltonian}) in the field theory limit. 
We will initially  set the on-site chemical potential $\epsilon_{d}=0$, but  reinstate it later.
First, let us  bosonize $\psi^{}_\pm=\frac{\eta_\pm}{\sqrt{2\pi}}\,e^{i\sqrt{4\pi}\varphi_\pm}$. 
This yields 
\be
\begin{array}{ccc}
H \! &=& \!
\sum_{a=\pm} H_0(\varphi_a) +
\frac{\gamma \kappa^+}{\sqrt{\pi}}\,\big[e^{i\sqrt{4\pi}\varphi_+}(0)\,S^+ +\mbox{h.c.}\big] 
\vspace*{0.1cm}\\
&& \hspace*{0.5cm}+ 
\frac{U}{\sqrt{\pi}}\,\left(\p_x\varphi_+(0) +\p_x\varphi_-(0)\right)\,S^z
\end{array}
\ee
where the free hamiltonian is  $H_0(\varphi_a) = \frac{1}{2}\int dx\,\left(\p_x\varphi_a\right)^2$ and $\kappa^+=\eta\eta_+$.
We then perform the unitary transformation  
$\Cal U=e^{i\alpha S^z \!\left(\varphi_+ \!+\!\varphi_-\right)(0)}$ 
and choose $\alpha=\frac{U}{\sqrt{\pi}}$, to cancel the remaining interaction along $S^z$. 
Define now
$\phi_+ = \frac{1}{\beta}\big[ (\sqrt{4\pi}\!-\!\alpha)\varphi_+ \!-\! \alpha\varphi_-   \big]$ 
and 
$\phi_- = \frac{1}{\beta}\big[ (\sqrt{4\pi}\!-\!\alpha)\varphi_- \!+\! \alpha\varphi_+   \big]$, 
with $\beta^2 = \frac{2}{\pi}(U - \pi)^2 + 2\pi$.
In terms of these new variables, the hamiltonian then becomes:
\be
H^I=\sum_{a=\pm }H_0(\phi_a) + \frac{\gamma \kappa_+}{\sqrt{\pi}}
\,\left[ e^{i\beta\phi_+}(0)\,S^+ 
+\mbox{h.c.} \right]
\label{hamI}
\ee
A remarkable feature is that the angle $\theta$ has disappeared from this expression; as a result, the free energy of the impurity problem  is independent of the anisotropy. 
Hamiltonian $H^{I}$ is formally equivalent to the anisotropic Kondo problem. 
The interaction term has scaling dimension $D=\frac{\beta^2}{8\pi}$ and thus we see immediately that in the scaling theory, tunneling is irrelevant in the low energy limit for $\beta^{2}>8\pi$ i.e.~strong Coulomb interaction $|U-\pi|>\sqrt{3}\pi$ and relevant otherwise. In the latter case, the $\phi_{+}$ hamiltonian flows  to the ordinary Kondo fixed point. We will soon argue this corresponds to the resonant level being hybridized with the wires, with a fixed, anisotropy dependent amount of tunneling between the two.

To proceed, observe that we could first  bosonize $\psi_{1,2}$ and then only form linear combinations, this time of the bosons.  
Setting $\psi^{}_a=\frac{\eta_a}{\sqrt{2\pi}}\,e^{i\sqrt{4\pi}\varphi_a}$ and forming the  combinations 
$\phi_1=\frac{1}{\sqrt{2}}\left(\varphi_1+\varphi_2\right)$, $\phi_2=\frac{1}{\sqrt{2}}\left(\varphi_1-\varphi_2\right)$, 
the Hamiltonian is then rotated by the \emph{same} unitary transformation
$\Cal U=e^{i\alpha\sqrt{2}\,S^z\,\phi_1(0)}$ 
to yield
\be
H^{\II} = \sum_{a=1,2} H_0(\phi_{a})  
+\frac{\gamma }{\sqrt{2\pi}}\,\left[\Cal V_1(0)\Cal O_2(0)
S^+ + \hbox{h.c.} \right]
\label{hamII}
\ee
where we have introduced the vertex operators
$\Cal{V}_{\pm 1} = e^{\pm i\beta_1\phi_1}$, 
$\Cal V_{\pm 2}  = e^{\pm i\sqrt{2\pi}\phi_2}$, 
$\Cal O_2= \gamma_{1}\kappa_1\Cal V_2+\gamma_{2}\kappa_2\Cal V_{-2}$ 
and $\kappa_a=\eta\eta_a$.
The parameter $\beta_1=\sqrt{2\pi}-\alpha\sqrt{2}$ satisfies $\beta^2=\beta^2_1+2\pi$ which ensures that the perturbations have the same scaling dimension $D$. 
This representation of the Hamiltonian is more suited to non equilibrium situation since the electrical current from wire 1 to 2 is simply expressible in terms of $\phi_2$ only.
However, $H^{\II}$ has a much more complex form than $H^{I}$, and typically mixes Kondo and boundary sine-Gordon (BSG) type interactions. This is particular clear in the case $\beta_{1}\!=\!0$: in the limit $\theta\to 0$, the Hamiltonian reduces to 
$H_{B}^{\II}= {\gamma\kappa_1 \over\sqrt{\pi}} e^{i\sqrt{2\pi}\phi_{2}}S^{+}+ \mbox{h.c.}$, a Kondo hamiltonian.
If, for the sake of argument, we neglect the Klein factors (this is allowed for example in the computation of the free energy), the case $\gamma_{1}=\gamma_{2}$ reduces to 
$H_{B}^{\II}= 4\gamma /\sqrt{2\pi}\cos(\sqrt{2\pi}\phi_{2})S^{y}$ 
i.e.~two copies of the BSG model for $S^{y}=\pm {1\over 2}$. We will see later that indeed, reintroducing the Klein factors, the model at $\beta_1=0$ and $\gamma_1=\gamma_2$ shares the same (Dirichlet) boundary conditions with the BSG model.

That the free energy of the two incarnations $H^{I}$ and $H^{\II}$ is the same and  independent of $\theta$ is remarkable. It can be  checked directly but not straightforwardly at all order of the  perturbative expansion in powers of $\gamma$.
 
We consider the question of the linear conductance at low energy, i.e.~when the resonant level is hybridized with the wires. A quick way to obtain it is to use the boundary conditions (BC's) for the fields $\phi_{1,2}$ in the IR. These are not so easy to obtain from hamiltonian $H^{\II}$. We will thus start from the BC's for the fields $\phi_{\pm}$, which are  known from the general analysis of the Kondo model.
The idea is to follow these BC's through the canonical transformations. This seems very hard due to the non linearities involved, but becomes possible once one recognizes the presence of  SU(2) affine currents (the SU(2) transformations are those mixing 
the two wires). Introducing
\be
\begin{array}{rclcrcl}
J^a &=& \frac{1}{2}\,:\psi^\dagger_\alpha\sigma^a_{\alpha\beta}\psi^{}_\beta:
& \! ,
& J^z &=& \frac {1}{\sqrt{2\pi}} \,\p_x \phi_2,\\
&&&&&& \vspace{-0.25cm}\\
J^x &=& \frac{-i\rho}{2\pi}\sin(\sqrt{8\pi}\phi_2) & \! ,
& \; J^y &=& \frac{-i\rho}{2\pi} \cos(\sqrt{8\pi}\phi_2),
\end{array}
\ee
(here $\rho=\kappa_1^{}\kappa_2^{}$) it is easy to show that 
\bea
\p_x\phi_- &=& 
\textstyle{\frac{\sqrt{2\pi}}{\beta}}
\; \p_x\phi_1  - \sqrt{2\pi} 
\textstyle{\frac{\beta_1}{\beta}} 
\left[ \cos\theta J^z + \sin\theta J^x\right]
\nonumber\\
\p_x\phi_+ &=& 
\textstyle{\frac{\beta_1}{\beta}}
\; \p_x\phi_1 + 
\textstyle{\frac{2\pi}{\beta}}
\, \left[ \cos\theta J^z + \sin\theta J^x\right]
\label{tfocanan}
\eea
We now recall that  in the IR the field $\phi_+$ obeys Neumann BC's with angle $\frac{\beta}{4}$, $\phi_{+}(0^+)=\phi_{+}(0^-)+{\beta\over 4}$,  and the boson $\phi_-$ -- being unaffected by the interaction -- Neumann BC's with angle $0$.
Introducing SU(2) rotated currents $\tilde J^a=\Cal R^y_{\theta}\cdot J^a$ ($\Cal R_{\theta}^y$ is a rotation of angle $\theta$ around $J^y$), we see that in the IR, these currents obey the BC's $\tilde J^z(0^+)=J^z(0^-)$ and $\tilde J^\pm(0^+)=-\tilde J^\pm(0^-)$.
A little algebra based on the SU(2) commutation relations then leads to 
\bea
J^z(0^+) &=& \cos(2\theta)\,J^z(0^-) + \sin(2\theta)\, J^x(0^-)\nonumber\\
J^x(0^+) &=& -\cos(2\theta)\,J^x(0^-) + \sin(2\theta)\, J^z(0^-)\nonumber\\
J^y(0^+) &=& -J^y(0^-)
\label{bcIRres}
\eea 
which of course are highly non linear in terms of the field $\phi_{2}$ itself. As for $\phi_{1}$, it obeys simply $\phi_{1}(0^+)=\phi_{1}(0^-)+{\beta_{1}\over 4}$. 
The BC's for  $\phi_2$ interpolate continuously between Neumann (N) 
($\theta\!=\!0$ and $\phi_2(0^+)\!=\! \phi_2(0^-)\!+\!\sqrt{\pi/8}$)
and double Dirichlet  (D)
($\theta\!=\!\frac{\pi}{2}$ and $\phi_2(0^+)\!=\! -\phi_2(0^-)\!\pm \!\sqrt{\pi/8}$). 
This is possible because the dimension of the operator $e^{\pm 
i\sqrt{2\pi}\phi_{2}}$ is the inverse of an integer square, here $\frac{1}{2^2}$ \cite{strings}. 
As a result, the ratio of boundary degeneracies for N and D is $g_{\mbox{\scriptsize N}}^{}/g_{\mbox{\scriptsize D}}^{}=2$, and degeneracies of 
N and double D are the same.
That the two fixed points can be reached depending on $\theta$ is particular clear in the case $\beta_{1}=0$ discussed previously, where $H^{\II}$ interpolates between Kondo and double BSG. Independently of $\theta$, the IR boundary degeneracy thus always takes the N value, $g_{\ir}=g_{\mbox{\scriptsize N}}^{}$.

Another way 
of viewing 
these BC's is to observe that the radius of compactification $r= \sqrt{2}$ being an integer multiple of the self dual radius $r^*=\frac{1}{\sqrt{2}}$ allows for the existence of a pair of non chiral operators of dimension one, $e^{\pm i\sqrt{2\pi}(\phi_{2}+\bar{\phi}_{2})}$, which induce an exactly marginal boundary deformation through $J^{x,y,z}$. To our knowledge, this is the first time such a BC is encountered in a condensed matter context. It is also the first time that a tunable  flow into a line of boundary fixed points is encountered.

A similar analysis can be carried out when there is a chemical potential for the electron on the dot, i.e.~a term $\epsilon_{d} d^{\dagger}d=\epsilon_{d}(S^{z}+\frac{1}{2})$. As before we first argue for the Kondo hamiltonian $H^{I}$.
The local magnetic field results in an additional phase shift for the electrons \cite{Natanold}, which translates into a phase shift $\delta_+$ for the field $\phi_{+}$. 
This is easy to understand: an additional phase shift for $\phi_{+}$ in the IR is tuned by the introduction of a scattering potential term, which then induces, by Friedel sum rule, an extra ``charge'' (here, magnetization) on the impurity.
Now, it is known that the problem with a field acting on the impurity only is closely related \cite{Lowenstein} (in the scaling limit) to the problem with a field coupling to the total spin $S_{tot}^{z}= S^{z}+{2\over \beta}\int \partial_{x}\phi_{+}$ which is  immediately solvable by BA. This gives rise to the impurity magnetization $m_{imp}=\vm{ S^{z}}=n_d-\frac{1}{2}$, with  the relation $\delta_{+}=\sqrt{\pi\over 2}\; m_{imp}$. At $T=0$, the impurity magnetization can be obtained using the Wiener Hopf technique. One finds two possible expansions. For small enough $\epsilon_{d}$ one has
\vspace*{-0.1cm}
 \begin{equation}
 \vspace*{-0.1cm}
m_{imp}={1\over D\sqrt{\pi}}
\sum_{n=0}^{\infty}{(-1)^{n}\over n!(2n+1)}
\frac{\Gamma\big(\frac{2n+1}{2(1-D)}\big)} {\Gamma\big(\frac{(2n+1)D}{2(1-D)}\big)}
u^{2n+1}
\end{equation}   
with 
$u=\frac{D}{\sqrt{\pi}} \frac{\Gamma(D/2(1-D))}{ \Gamma(1/2(1-D))} \frac{\epsilon_{d}}	{T_{B}}$.
The dual expression should be used beyond the radius of convergence $u^*=D^{\frac{D}{2(1-D)}}\sqrt{1\!-\!D}$:
\vspace*{-0.1cm} 
\begin{equation}
\vspace*{-0.1cm}
m_{imp}={1\over 2\sqrt{\pi}}
\sum_{n=0}^{\infty}{(-1)^{n}\over n!}{\Gamma({1\over 2}\!+n(1-D))\over
\Gamma(1-nD)}u^{2n(D-1)}
\end{equation}
The parameter $T_{B}$ is the Kondo temperature for the problem defined by hamiltonian $H^{I}$, related to the bare coupling by 
$T_{B}/W = \frac{ \Gamma(D/2(1-D)) \Gamma(1-D)^{\frac{1}{1-D}}} {\sqrt{\pi}\; \;\Gamma(1/2(1-D))} \big(\frac{\gamma}{\sqrt{W}}\big)^{\frac{1}{1-D}}$, 
with $W$ the bandwidth (this relation holds within the regularization inherited from  integrability used in Ref.\cite{LS}). 
 
The additional phase shift for $\phi_{+}$ translates into more complicated non linear BC's for the boson $\phi_{2}$: namely, the rotated currents $\tilde J^a$ now obey the BC's $\tilde J^z(0^+)=\tilde J^z(0^-)$ and $\tilde J^\pm(0^+)=-e^{\pm i\Delta}\tilde J^\pm(0^-)$, with the angle $\Delta=2\pi m_{imp}$. After some calculations one finds 
\be
\begin{array}{l}
J^{z}(0^{+})=\left[\cos^{2}\theta-\sin^{2}\theta\cos\Delta\right]J^{z}(0^{-})
\vspace{0.1cm}\\
\hspace{0.6cm}+\sin(2\theta)\cos^2\frac{\Delta}{2}\;J^{x}(0^{-})+\sin \theta\sin\Delta 
\;J^{y}(0^{-})\hphantom{\hspace{0.65cm}}\vspace{0.2cm}
\end{array}
\nonumber
\vspace*{-0.65cm}
\ee
\be
\begin{array}{l}
J^{x}(0^{+})=\left[\sin^{2}\theta-\cos^{2}\theta\cos\Delta\right]J^{x}(0^{-})
\vspace{0.1cm}\\
\hspace{0.6cm}+\sin(2\theta)\cos^2\frac{\Delta}{2}\;J^{z}(0^{-})-\cos \theta\sin\Delta 
\;J^{y}(0^{-})\vspace{0.2cm}
\end{array}
\vspace*{-0.65cm}
\label{bcIRnonres}
\ee
\be
\begin{array}{l}
J^{y}(0^{+})=-\sin \theta\sin\Delta \;J^{z}(0^{-})\vspace{0.1cm}\\
\hspace{0.6cm}+\cos \theta\sin\Delta \;J^{x}(0^{-})-\cos\Delta \;J^{y}(0^{-}) 
\hphantom{\hspace{1.9cm}}
\end{array}
\nonumber
\ee
These BC's relate the currents on both sides of the impurity through a SU(2) rotation now depending on the anisotropy and doping of the impurity.
     
To extract information from the BC's in the IR, it is convenient to reformulate them first within a boundary field theory by folding and introducing complex coordinates  $z=\tau-ix$, $x>0$. The Kubo formula then reads:
\be
G\!=\!
\lim_{\omega\to 0}\frac{e^2}{\hbar} 
\frac{1}{(2L)^2}\frac{1}{\omega}\!\int\! dx dy d\tau e^{i\omega \tau} 
\vm{ j_e(x,\tau)j_e(y,0)}
\label{kubo}
\ee
where $0<\tau<T^{-1}$, spacial integrals run over $[0,L]$ and $L$ has to be sent to $+\infty$;  the electrical current through the whole structure is 
$j_e(x) = 2\big(J^z(x)-J^z(-x)\big)$ 
(the conductance depends only on the $\phi_2$ propagator).

Using the propagators that can be deduced from (\ref{bcIRnonres}):
\bea
4\pi\vm{\p_x \phi_2(z)\p_x\phi_2(w)} &=& 
\prt{z-w}^{-2}
\label{propPhiIR}\\
4\pi\vm{\p_x \phi_2(z)\p_x\phi_2(w^*)} &=& 
 \big(1-2\sin^2\theta\cos^2\!\textstyle\frac{\Delta}{2}\big) \prt{z-w^*}^{-2} 
\nonumber
\eea
one finds 
$G_\ir =\frac{e^2}{h}\, \sin^2\theta\cos^2\frac{\Delta}{2}$, 
which is nothing but Eq.(\ref{gfriedel}). The capacitive coupling $U$ has disappeared at the IR fixed point: it is 'irrelevant', but as we will see, it controls the approach of the fixed point and thus determines the low $T<T_B$ properties of the theory, which we will now be able to tackle, thanks to this long reformulation of Friedel sum rule.

Indeed, the exact solution of the Kondo hamiltonian leads to a full knowledge of the infinity of  counter-terms necessary to describe the approach to this fixed point \cite{LS}, allowing one to carry out IR perturbation theory to all orders. A program such as the one of \cite{LSi} could then lead to results for the linear conductance at arbitrary values of the temperature. 
It relies on the identification of the low $T$ Hamiltonian, which has the form:
$H=H_\ir + \sum_{k>0} b_{2k-1}\Cal O_{2k}(x=0)$.
It is important to stress that this expansion is highly non perturbative in the tunneling amplitude (as we will see below, it leads to an expansion of the conductance in powers of $\gamma^{-2/1-D}$).
But the couplings $b_{2k-1}$ turn out to be  known explicitly \cite{LSi}! The whole set of perturbing operators $\Cal O_{2k}$ is a set of commuting conserved quantities related to integrability and describes the approach to the IR fixed point; it is made of fields of even dimensions. $\Cal O_{2k}$ can be expressed as a polynomial in $\p_x\phi_+$ and its derivatives, to be then translated in the $\phi_{1,2}$ basis. We just sketch here the (somewhat lenghty) analysis. Apart from density-density couplings, the leading irrelevant contribution contains a tunneling term 
$\Cal O_2^{\mbox{\scriptsize tun}}=\lambda 
 (\psi_1^{\dagger}\psi^{}_{1}+\psi_{2}^{\dagger}\psi^{}_{2})
\big(\psi_{1}^{\dagger}\psi^{}_{2}+\hbox{h.c.}\big)$
with a coupling constant $\lambda\propto {\beta_{1}\over \beta^2}\sin\theta$. 
The anisotropy and coulombic repulsion are only apparent in the amplitude of the tunneling term -- which, as it should, vanishes in the Kondo limit, $\theta=0$. This pattern generalizes to all orders: the whole set of operators describing the approach to the IR fixed point is independent of $U$.

\begin{figure}
\includegraphics[angle=0,width=7cm]{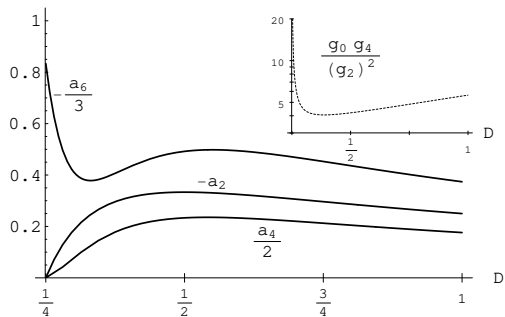}
\vspace*{-0.45cm}
\caption{First three reduced coefficients $a_{2k}^{}=g_{2k}^{}(T_B/\pi)^{2k}$ (defined in (\ref{girpert})).  The scaling dimension $D$ varies between $\frac{1}{4}$ and 1, which corresponds to the region $|U/\pi-1| <\sqrt{3}$ where tunelling is relevant.
The inset displays the universal ratio $\rho=g_0^{}g_4^{} / g_2^2$, which diverges for  $D=\frac{1}{4}$ ($U=\pi$).}
\label{corrNeB}
\vspace*{-0.4cm}
\end{figure} 

To obtain the conductance, the current-current correlator in Eq. (\ref{kubo}) is expanded in powers of the couplings $b_{2k-1}$. The resulting multiple integrals over intermediate times of finite $T$ correlators are evaluated using the residue theorem; divergences are regularized in the 'integrable' scheme through the commutativity of the $\Cal O_{2k}$. 
This way, we extract the low $T$ expansion of $G$, yielding the 'Landau-Fermi parameters' $g_{2k}$ for the conductance:
\be
G=G_\ir\;\big(1+\sum_{k>0} g_{2k}^{}T^{2k}\big).
\label{girpert}
\vspace{-0.2cm}
\ee
It is important to realize that the coefficients in this expansion are \emph{universal} in the field theory limit. They can be put in the form $g_{2k}^{}=a_{2k}^{}\big(\frac{\pi}{T_B}\big)^{2k}$ with $a_{2k}^{}$ depending only on $U$. Laborious calculations yield:
\be
a_2 = - \frac{4X}{3(1+X)^2}
\hphantom{\hspace*{4.16cm}}
\label{corrg}
\ee
\vspace*{-0.7cm}
\be
a_4 =
\frac{16X}{45(1+X)^2}\left[1+\frac{3X}{(1+X)^2} + y\frac{X(15-X)}{16\pi}\right]
\nonumber
\ee
where we introduced the parameter $X=4D-1$, and 
$y=\frac{\Gamma\left( D/2(1-D)\right)^{3}    \Gamma\left( 3/2(1-D) \right)}
{\Gamma\left(3D/2(1-D)\right) \Gamma\left( 1/2(1-D) \right)^{3}}$. 
On fig.\ref{corrNeB}, the $a_{2k}$'s are plotted up to order $2k=6$ (we have obtained $a_6$ as well, but its expression is  too lengthy to be shown here).
 
 The lowest order correction $g_2^{}$ can be understood in a simple way: at generic values of the coulombic repulsion on the impurity, the IR fixed point is a Fermi liquid whose approach is controlled at lowest order by a single operator, namely the energy momentum tensor $(\p\phi_+)^2$. Now corrections to $G_{\ir}$ can only stem from this part $\Cal O_2^{\mbox{\scriptsize tun}}$ of the perturbing operator that involve charge transfer across the impurity site, whose amplitude is simply multiplicatively renormalized with respect to the free case: $\lambda(U)=\frac{2\sqrt{X}}{1+X}\lambda(0)$.
This reasoning fails for higher orders, that are controlled by several processes with different  couplings, each of them being a function of $U$.

Note that the $U$-dependance of $T_B$ results in a maximum for $G$ in the scaling limit $\gamma/\sqrt{W}\ll 1$ ; this effect, which comes from the boundary perturbation being most relevant for $D=\frac{1}{4}$ (corresponding to $U=\pi$ in our renormalization scheme), was also noted in \cite{Borda} at $T=0,V\neq 0$  using perturbation theory in $U$.

The Landau-Fermi parameters allow to form a number of universal ratio, the simplest one being $\rho=\frac{g_0^{}g_4^{}}{g_2^2}=\frac{a_4^{}}{a_2^2}$. The first order of its development in $U$ agrees with results in \cite{Benjamin}. It displays (see fig.1) a divergence at the particular value $U=\pi$ of the Coulombic repulsion, which might offer an efficient way to identify this point in experimental realizations of the IRLM. 
Moreover, at this value of $U$, there are \emph{no} processes allowing for charge transfer up to order 6: the coefficients $a_2$ and $a_4$ vanish -- indicating the somewhat singular nature, for transport properties, of the Fermi liquid at this point -- while $a_6\big|_{U=\pi}\!=-\frac{1}{105}\big(\frac{2\pi}{\Gamma(2/3)^3}\big)^6$. This results in a further enhancement of the conductance around $U=\pi$ (see fig.2).

\begin{figure}[h]
\vspace*{-0.3cm}
\includegraphics[angle=0,width=7cm]{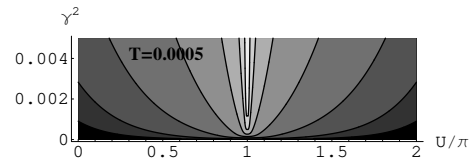}
\vspace*{-0.35cm}
\caption{Iso-conductance plot in the $(U/W,\gamma^2/W)$ plane, at fixed T/W=0.0005. On each line, $G/G_\ir=1-10^{-x}$ with $x$ ranging from 1 (dark) to 7 (bright).}
\label{isoG}
\vspace*{-0.25cm}
\end{figure}

In conclusion, it should be clear that methods of field theory give one a complete control of the linear conductance problem  from the IR point of view. Apart from their practical use (the $8^{th}$ order could be calculated and the series Pade resummed to obtain full crossover curves)  we hope that our results will provide useful benchmarks in assessing other approaches to the problem. 

The authors would like to thank B. Doyon  and A. Komnik for numerous enlightening discussions,  as well as  N. Andrei and P. Mehta for their patient explanations. 


\end{document}